\documentclass[final,letterpaper]{aipproc}
\layoutstyle{8x11double}
\usepackage{xspace}

\newcommand{\Ep}[1]{{\ensuremath{10^{#1}}}}

\newcommand{\mathff}[1]{\mathrm{#1}}

\newcommand{\Msun}{{\ensuremath{\mathff{M}_{\odot}}}\xspace}

\newcommand{\Sec}{{\ensuremath{\mathff{s}}}\xspace}
\newcommand{\km}{{\ensuremath{\mathff{km}}}\xspace}
\newcommand{\kms}{{\ensuremath{\km\,\Sec^{-1}}}\xspace}

\newcommand{\ms}{{\ensuremath{\mathff{ms}}}\xspace}
\newcommand{\cm}{{\ensuremath{\mathff{cm}}}\xspace}
\newcommand{\g}{{\ensuremath{\mathff{g}}}\xspace}
\newcommand{\gcc}{{\ensuremath{\g\,\cm^{-3}}}\xspace}

\newcommand{\aBH}{{\ensuremath{a_{\mathff{BH}}}}\xspace}

\newcommand{\FIG}[1]{\ref{fig:#1}}
\newcommand{\Figff}[1]{\FIG{#1}\xspace}
\newcommand{\Fig}[1]{Figure~\FIG{#1}\xspace}
\newcommand{\Figs}[1]{Figures~\FIG{#1}\xspace}

\begin{document}

\title[Progenitors of Collapsars]{On the Progenitors of Collapsars}


\author{A. Heger}{
  address={Department of Astronomy and Astrophysics, University of California, Santa Cruz, CA 95064, USA},
  email={alex@ucolick.org},
  homepage={http://2sn.org},
  altaddress={now at}{Department of Astronomy and Astrophysics, University of Chicago, Chicago, IL 60637, USA}
}

\author{S. E. Woosley}{
  address={Department of Astronomy and Astrophysics, University of California, Santa Cruz, CA 95064, USA},
  email={woosley@ucolick.org},
  homepage={http://supersci.org}
}

\copyrightyear  {2001}

\begin{abstract}
We study the evolution of stars that may be the progenitors of common
(long-soft) GRBs. Bare rotating helium stars, presumed to have lost
their envelopes due to winds or companions, are followed from central
helium ignition to iron core collapse. Including realistic estimates
of angular momentum transport \citep{HLW00} by non-magnetic processes
and mass loss, one is still able to create a collapsed object at the
end with sufficient angular momentum to form a centrifugally supported
disk, i.e., to drive a collapsar engine.  However, inclusion of
current estimates of magnetic torques \citep{Spr01} results in too
little angular momentum for collapsars.
\end{abstract}

\date{\today}

\maketitle

\section{Introduction}

One of the most promising models for the ``long variety'' of gamma-ray
bursts (GRBs) is the so-called \emph{collapsar} model \citep{Woo93}.
It assumes that a sufficiently massive stellar core collapses into a
black hole and the infalling outer layers form a disk around it.
Energy dissipated in the disk or the rotation of the black hole itself
is assumed to power a jet of high Lorentz factor ($\Gamma\sim200$) that
escapes from the engine along the polar axis to large distance
($\sim\Ep{15}\,\cm$) and powers a GRB by interaction with the
circumstellar medium or by internal shocks.

The traversal time for the relativistic jet through the hydrogen
envelopes of typical massive stars is hundreds to thousands of
seconds.  Thus, at the time of the GRB, bare helium stars, which have
radii of only a few light seconds (about a solar radius), are required
if the lifetime of the engine and the GRB are not to be short compared
to the time it takes the jet to drill through the star.

Two essential ingredients for the collapsar model are a sufficiently
massive core to form a black hole and a sufficient rotation rate at
the time of collapse to allow the formation of a disk.  The question
we address here is: \emph{What can be expected for the rotation rates
of massive stellar cores when they collapse?}

\section{Progenitor Models}

We calculate the evolution of bare helium cores with an initial mass
of 15\,\Msun.  The stars are assumed to rotate rigidly initially with
two different surface rotation rates corresponding to 10\,\% and
30\,\% of a Keplerian orbit.  The former could be either the result of
a massive single star ($\sim40\,\Msun$) that has lost its envelope
early during helium burning or a close binary that lost its envelope
to a companion.  The latter might require a binary merger.  The
evolution of the helium core and its rotation is followed as described
by \citet{HLW00} using fine surface zoning.

Two different evolutionary paths are considered.  The first neglects
mass loss, possibly corresponding to WR stars of very low metallicity,
while in the second, it is taken into account.  We use the WR mass
loss rate given by \citet[equation 1]{WL99}, reduced by a factor 3 to
account for effect of ``clumping'' \citep{HK98}.  An initial stellar
metallicity of 1/10 solar is assumed along with a WR mass loss rate
that scales as the square root of metallicity \citep{Van01}, reducing
the mass loss rate by an additional factor of 3.

\section{Results}

\begin{figure}
\includegraphics[width=\columnwidth]{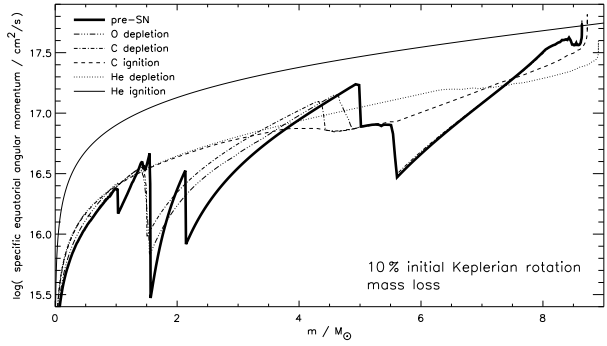}
\caption{Angular momentum in the equatorial plane as a function of the
interior mass coordinate, $m$, at different evolutionary stages.
\label{fig:b3jj}}
\end{figure}

\Figs{b3jj} - \Figff{g3j} give the results for collapsar progenitors
that follow the evolution of the angular momentum in the stellar
interior till the onset of iron core collapse.

\subsection{Rotation Profile and Disk Formation}

\begin{figure}
\includegraphics[width=\columnwidth]{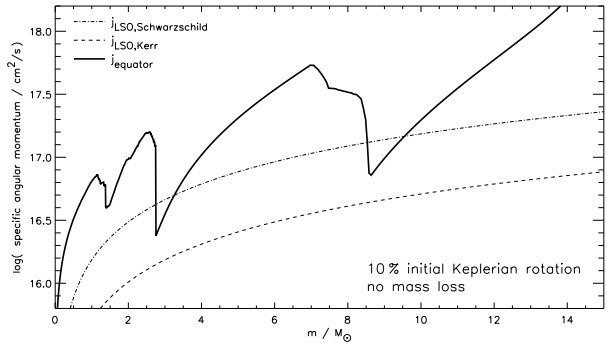}
\caption{Equatorial angular momentum (\textsl{thick black line}) of a
15\,\Msun helium core of initially 10\,\% Keplerian rotation as a
function of the interior mass coordinate, $m$.  The
\textsl{dashed-dotted line} shows the specific angular momentum
required for a test particle at the last stable orbit around a
Schwarzschild black hole of mass equal to the mass coordinate, $m$.
The \textsl{dashed line} shows the same for a extreme Kerr black hole
(spin parameter $a=1$). }
\label{fig:a2j}
\end{figure}

\begin{figure}
\includegraphics[width=\columnwidth]{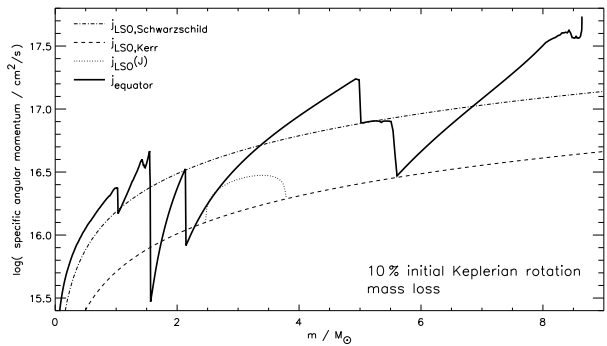}
\caption{Same as \Fig{a2j}, but mass loss due to stellar winds is
included.  The \textsl{dotted line} shows the specific angular
momentum a test particle requires at the last stable orbit around a
black hole that has formed with the mass and integrated angular
momentum below the given mass coordinate.  Where the dotted line is
missing, sufficient angular momentum is available to form a Kerr black
hole.}
\label{fig:b3j}
\end{figure}

To decide whether a centrifugally supported accretion disk can form
around a central black hole (which we assume either forms promptly or
by fallback), we compare the angular momentum calculated as a function
of interior mass to that a test particle would require at the last
stable orbit around a Schwarzschild or Kerr black hole (\Figs{a2j} and
\Figff{b3j}).  Though mass loss significantly reduces the angular
momentum at core collapse, enough remains in the equatorial regions of
the star to form a centrifugally supported disk.


\subsection{Magnetic Fields}

\begin{figure}
\includegraphics[width=\columnwidth]{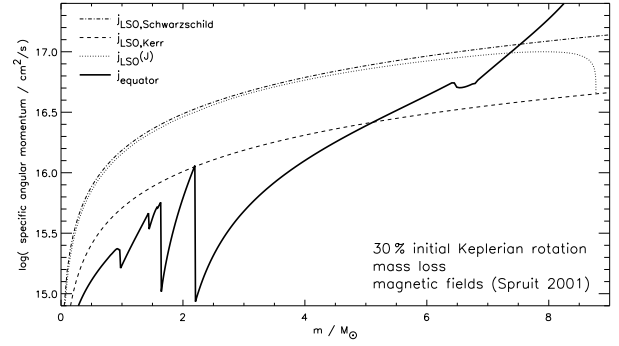}
\caption{ Same as \Fig{b3j}, but an initial rotation rate of 30\,\%
Keplerian rotation was assumed and angular momentum transport by
magnetic fields according to a prescription by \citet{Spr01} was
included.}
\label{fig:g3j}
\end{figure}

We also calculated models including magnetic torques that might result
from a dynamo process recently described by \citet{Spr01}.  They lead
to considerable spin-down of the core, especially when combined with
wind mass loss (Table~\ref{tab:bh} and \Fig{g3j}).

For comparison, preliminary calculations \citep{HWS02} of rotating
stellar models of 15, 20, and 25\,\Msun stars with hydrogen envelopes
and initial surface rotation velocities of 200\,\kms, using the same
prescription for torques, resulted in pulsar birth rotation periods of
7.65, 5.50, and 3.99\,\ms.  Without magnetic fields, including all
other forms of angular momentum transport \citep{HLW00}, the same
calculations previously gave rotation periods of $\sim$0.2\,\ms in
these same stars.  Here it is assumed that the innermost 1.6\,\Msun of
the collapsing star produces a rigidly rotating neutron star of
1.45\,\Msun gravitational mass, while conserving the total angular
momentum contained in the pre-collapse model.

\begin{table}
\scalebox{0.975}{
\begin{tabular*}{\columnwidth}{@{}c@{\extracolsep{\fill}}c@{\extracolsep{\fill}}c|@{}c|@{}c@{\extracolsep{\fill}}c@{\extracolsep{\fill}}c@{\extracolsep{\fill}}c|@{}c@{}}
\rotatebox{90}{magnetic fields} &
\rotatebox{90}{rotation}\phantom{l}\rotatebox{90}{(\% Keplerian)} &
\rotatebox{90}{mass loss} &
\rotatebox{90}{pulsar period}\phantom{l}\rotatebox{90}{(\ms)} &
\rotatebox{90}{\aBH (2\,\Msun)} &
\rotatebox{90}{\aBH (2.5\,\Msun)} &
\rotatebox{90}{\aBH (3\,\Msun)} &
\rotatebox{90}{\aBH (4\,\Msun)} &
\rotatebox{90}{mass coordinate}\phantom{l}
\rotatebox{90}{ranges for which}\phantom{l}
\rotatebox{90}{an equatorial}\phantom{l}
\rotatebox{90}{disk could form}\phantom{l}
\rotatebox{90}{(\Msun)} 
\\
\hline
 -- & 10 &  -- & 0.09 & (2.5) & (3.4) & (3.8) & (3.7) & $ 0-15$ \\
 -- & 10 & yes & 0.23 & (1.1) &  0.90 &  0.98 & (1.2) & $ 0- 2.5$, $ 2.7- 8.6$ \\
 -- & 30 &  -- & 0.06 & (4.1) & (4.8) & (5.6) & (5.9) & $ 0-15$ \\
 -- & 30 & yes & 0.18 & (1.7) & (1.3) & (1.3) & (1.7) & $ 0- 8.4$ \\
\hline           
yes & 10 &  -- & 0.45 &  0.58 &  0.76 &  0.80 &  0.88 & $ 2.4- 2.8$, $ 5.7-15$ \\
yes & 10 & yes &  3.4 &  0.08 &  0.09 &  0.11 &  0.10 & $ 8.5- 8.8$ \\
yes & 30 &  -- & 0.26 & (1.2) & (1.4) & (1.4) & (1.5) & $ 0-15$ \\
yes & 30 & yes &  1.9 &  0.16 &  0.14 &  0.16 &  0.17 & $ 7.4- 8.8$ \\
\hline           
\end{tabular*}
} 
\caption{Presupernova properties of initially 15\,\Msun helium core
models.  The first three columns define the initial model and physics
employed (magnetic fields according to \citet{Spr01}, amount of
rotation, mass loss by winds).  Next we give the period a pulsar would
have if it formed in this star, then the non-dimensional spin
parameter, \aBH, a black hole would acquire, if all the angular
momentum below the mass coordinate indicated were to go into the black
hole of that mass (formal values in excess of 1 are shown in brackets
solely to give a measure for the angular momentum available in the
model), and in the last column we show the mass ranges in which the
equatorial mass could form a centrifugally supported accretion disk
around a central compact object.}
\label{tab:bh}
\end{table}

Calculations that used the effects of magnetic fields as described by
\citet[not shown here]{SP98} always resulted in much too low core
rotation for the collapsar model of GRBs.

\subsection{Binary Interaction}




In a sufficiently close binary system the envelope can be tidally
locked to the orbit of the star.  Depending on separation and mass
ratio, this rotation can reach several 10\,\% of Keplerian.  We find
that maintaining 10\,\% Keplerian surface rotation can be sufficient
for collapsars.  If employing the dynamo process by \citet{Spr01},
however, the resulting presupernova rotation is too slow even for a
system with 20\,\% Keplerian surface rotation.

To simulate a merger of a binary system after or at the end of central
helium burning (Case~C) an additional model was calculated assuming
rigid rotation with 50\,\% Keplerian surface rotation at core helium
depletion and including the dynamo process of \citet{Spr01}.  Again the
magnetic stress kept the star in rigid rotation till carbon burning,
removing too much angular momentum for the collapsar model to work.
\emph{Without} magnetic fields, as shown above, even single stars may
already retain sufficient angular momentum for collapsar progenitors.

\subsection{Decoupling of Core Rotation}

\begin{figure}
\includegraphics[width=\columnwidth]{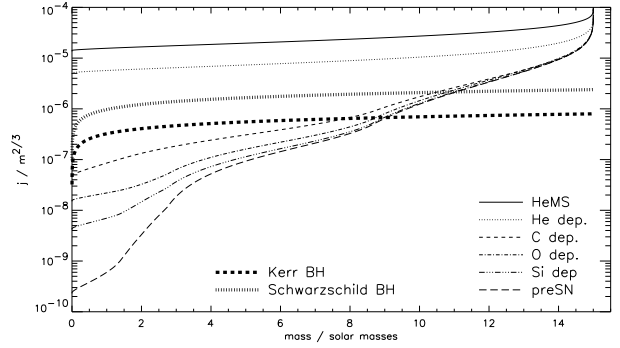}
\caption{Angle averaged specific angular momentum for a rigidly
rotating 15\,\Msun star with Keplerian surface rotation at different
evolution stages.  The specific angular momentum has been scaled by
$m^{-2/3}$ to remove the singularity at $m=0$ from the plot.
\textsl{Thick lines} show the specific angular momentum required for
the last stable orbit around a black hole.}
\label{fig:x2c}
\end{figure}

\emph{At what evolution stage does the core rotation need to decouple
from the surface?}\/ \Fig{x2c} shoes that this needs to happen before
central carbon burning -- assuming angular momentum is locally
conserved from this time on and no further transport occurs.  In the
model star of \Fig{x2c} the decoupling would need to happen no later
than when a central density of $\sim$10,000\,\gcc is reached.

\section{Conclusions}

A bare helium star of low metallicity can retain enough angular
momentum to form a centrifugally supported disk around a central black
hole of $\sim3\,\Msun$, as required by the collapsar model for GRBs.
Without magnetic fields, the angular momentum is sufficient to form a
Kerr black hole and support most or all of the star in an accretion
disk.  However, if we include an approximate treatment of angular
momentum transport by magnetic fields, the resulting spin rates become
too low to form centrifugally supported disks in the inner part of the
core. Even a binary helium star merger at the end of central helium
burning might not be able to avoid this fate.  Mass loss can lead to
an additional significant spin-down of the core, especially if
magnetic fields couple it effectively to the envelope.  Even in case
of Keplerian surface rotation, the core rotation needs to decouple
before carbon ignition in order to make a Kerr black hole.  The dynamo
process recently proposed by \citet{Spr01} seems too efficient to form
collapsar progenitors from single stars or helium star mergers.  This
is even more so for the magnetic field modeling suggested by
\citet{SP98}.

\begin{theacknowledgments}
We thank Henk Spruit for a preview of his work and many helpful
discussions.  This work has been supported by the NSF (AST-9731569),
NASA (NAG5-8128), the DOE (B347885), and the AvH (FLF-1065004).
\end{theacknowledgments}
\vspace{-0.1in}


\end{document}